\documentclass[doublecol]{epl2}
\usepackage{color}
\usepackage{txfonts}
\usepackage{epsfig}
\usepackage{graphicx}
\usepackage{dcolumn}
\usepackage{bm}
\title{The emergent longitudinal wave from space and time derivatives transformations }
\shorttitle{The emergent longitudinal wave from space and time derivatives transformations}
\author{A. I. Arbab\inst{}\footnote{aiarbab@uofk.edu}}

\institute{
  \inst{} Department of Physics,
Faculty of Science, University of Khartoum, P.O. Box 321, Khartoum
11115, Sudan}


\abstract{
We have shown that a longitudinal wave emerges as a result of  general transformations similar to gauge transformations of electrodynamics.
The time derivative and the gradient of the gauge function and their alike yield the longitudinal wave. The de Broglie wave associated with the particle motion is a longitudinal one. This wave accompanied the motion of all massive objects. It has a zero magnetic field (or vorticity). The invariance of Dirac equation under these transformations makes the electron to behave as a massless particle while having a mass.}

\begin{document}

\maketitle

\section{Introduction}
De Broglie hypothesize that a particle has a wave nature associated with its motion. However, this wave nature is not determined by de Broglie.
Sch\"{o}dinger avail this property and constructed his wave theory of particles. In Sch\"{o}dinger formulation a plane wave is considered. But a particle is supposed to be described by a wavepacket.  Hence, a particle will then be assumed to be the superposition of some plane waves. Arbab has recently developed a wave theory describing the spin -1/2 and spin -0 particles ~\cite{arbab}. Unlike the ordinary firs-order space and time Dirac equation, Arbab expressed Dirac equation in a second-order wave equation form. Thus,  Klein-Gordon and Dirac equations become equally represented.
The reliable equation of motion has to be invariant under Lorentz transformations. However, Maxwell equations exhibit further gauge invariance. Owing to Nother's theorem, an invariance of the equation of motion under some transformation ushers into a conserved physical quantity. Unfortunately, the presence of the mass term in the equation of motion will spoil the gauge invariance of Maxwell equations. For this reason the photon (gauge field) is taken to have a zero mass. In electroweak theory the  mass of the particles, besides the gauge fields, were taken to be zero, and given their masses by the Higgs mechanism when the electroweak symmetry is broken ~\cite{higg}.

We propose here energy and momentum transformations, i.e., space and time derivatives transformations, and dictate that the equations of motion should be invariant under them. These transformations have some links with gauge transformations.
If we demand that the transformations of  space and time derivatives leave physical world unaltered, a longitudinal wave will result representing the material nature of the particle under consideration. In electrodynamics the  wave nature of the moving charged particle has not been investigated before. Maxwell equations describe the propagation of the electric and magnetic fields, without taking into account the material wave nature of the charged particle. The fields have a transverse wave  nature while the particle has a longitudinal wave nature \cite{arbab2}.

We will investigate in this letter the space and time derivative transformations (energy and momentum)  and their relation to the dynamics of the moving particle (charged or not). While the wave associated with the field is transverse, the wave associated with the particle is longitudinal. The magnetic field-like vanishes for longitudinal wave.
\section{General derivatives transformations}
Consider the following transformations
\begin{equation}
\vec{\nabla}\,'=\vec{\nabla}+\vec{\nabla}f\,,\qquad \frac{\partial}{\partial t}\,'=\frac{\partial}{\partial t}-\frac{\partial f}{\partial t}\,,
\end{equation}
where $f$ is a scalar (gauge) function.
\subsection{Maxwell equations}
If we apply eq.(1) in Maxwell equations ~\cite{girr},
\begin{equation}
\vec{\nabla}\times\vec{E}=-\frac{\partial\vec{B}}{\partial t}\,,\qquad\qquad  \vec{\nabla}\cdot\vec{E}=\frac{\rho}{\varepsilon_0}\,,
\end{equation}
and
\begin{equation}
\vec{\nabla}\cdot\vec{B}=0\,,\qquad \vec{\nabla}\times\vec{B}=\mu_0\vec{J}+\frac{1}{c^2}\frac{\partial\vec{E}}{\partial t}\,,
\end{equation}
we will obtain
\begin{equation}
\vec{\nabla}f\cdot\vec{B}=0\,,\qquad \vec{\nabla}f\cdot\vec{E}=0\,,
\end{equation}
and
\begin{equation}
(\vec{\nabla}f)\times\vec{E}=\frac{\partial f}{\partial t}\vec{B}\,,\qquad (\vec{\nabla}f)\times\vec{B}=-\frac{1}{c^2}\frac{\partial f}{\partial t}\vec{E}\,.
\end{equation}
The electric and magnetic fields created by a moving  charged particle satisfy, then $\vec{v}\cdot\vec{E}=\vec{v}\cdot\vec{B}=0$. Thus, it seems that $\vec{\nabla}f\propto\vec{v}$. This means that $\vec{\nabla}f$ is directed parallel to the motion of the particle. It thus leads to a generation of longitudinal wave.
Now if
\begin{equation}
\vec{\nabla}f=\frac{\vec{v}}{c^2}\frac{\partial f}{\partial t}\,,
\end{equation}
then eq.(5) represents the vanishing electric and magnetic fields in the boost frame ($\vec{E}\,'=0$ and $\vec{B}\,'=0$) with speed $\vec{v}$, i.e.,
\begin{equation}
\vec{B}=\frac{\vec{v}}{c^2}\times\vec{E}\,,\qquad \vec{E}=-\vec{v}\times\vec{B}\,.
\end{equation}
In quantum mechanics, $p=-i\hbar\vec{\nabla}$, $E=i\hbar\frac{\partial}{\partial t}$, so that eq.(6) yields the Einstein energy equation
\begin{equation}
p=\frac{v}{c^2}E\,\qquad\Rightarrow \qquad E=mc^2\,.
\end{equation}
Moreover,  eqs.(4) and (5) yield
\begin{equation}
\frac{1}{c^2}\frac{\partial^2f}{\partial t^2}-\nabla^2f=0\,.
\end{equation}
 Therefore, the invariance of Maxwell equations under the transformations given by eq.(1) is equivalent in electrodynamics to Lorenz gauge condition invariance under gauge transformations.

The effect of the interaction of an  electromagnetic field on the dynamic of a charged free particle is to replace its  momentum and energy by
\begin{equation}
\vec{p}\,'=\vec{p}+q\vec{A}\,,\qquad E\,'=E+q\varphi\,.
\end{equation}
In quantum mechanics, $p=-i\hbar\vec{\nabla}$, $E=i\hbar\frac{\partial}{\partial t}$. Hence, eq.(6) becomes
\begin{equation}
\vec{\nabla}\,'=\vec{\nabla}+i\frac{q}{\hbar}\vec{A}\,,\qquad \frac{\partial}{\partial t}\,'=\frac{\partial}{\partial t}-i\frac{q}{\hbar}\varphi\,.
\end{equation}
The ordinary gauge transformations are written as
\begin{equation}
\vec{A}\,'=\vec{A}+\vec{\nabla}f\,,\qquad \varphi\,'=\varphi-\frac{\partial f}{\partial t}\,.
\end{equation}
The Lorenz gauge condition requires that the gauge function $f$ to satisfy the wave equation
\begin{equation}
\frac{1}{c^2}\frac{\partial^2f}{\partial t^2}-\nabla^2f=0\,.
\end{equation}
The term $\vec{\nabla}f$ in eq.(6) can be considered as a longitudinal component of the vector field $\vec{A}$. Hence, the gauge transformation in eq.(6) is equivalent to decomposing the gauge field into transverse ($\vec{A}_\bot$) and longitudinal ($\vec{A}_\|$) components, i.e., $\vec{A}=\vec{A}_\bot+\vec{A}_\|$, \emph{viz}., $\vec{A}_\bot=\vec{A}$ and $\vec{A}_\|=\vec{\nabla}f$. Hence, the transformations in eq.(11) \& (12) and eq.(1) are intimately related.

\subsection{Dirac equation}
Let us now apply the transformations given by  eq.(1) in Dirac equation ~\cite{drell}
\begin{equation}
\frac{1}{c}\frac{\partial\psi}{\partial t}+\vec{\alpha}\cdot\vec{\nabla}\psi+\frac{imc\beta}{\hbar}\psi=0\,,
\end{equation}
to obtain
\begin{equation}
-\frac{1}{c}\frac{\partial f}{\partial t}+\vec{\alpha}\cdot\vec{\nabla}f=0\,.
\end{equation}
Applying eq.(6) in eq.(15) yields $\vec{\alpha}\cdot\vec{v}=c$. Using the fact that $\alpha^2=1$, this implies that $\vec{v}=\vec{\alpha}\, c$.
Interestingly, this is the result obtained in Dirac's formulation for the electron. Differentiate both sides of eq.(15) partially with respect to time, and use eq.(15) once again or eq.(6) once again to obtain
\begin{equation}
\frac{1}{c^2}\frac{\partial^2f}{\partial t^2}-\nabla^2f=0\,.
\end{equation}
 Equation (16) is the same as eqs.(9) \& (13). Hence, the invariance of Dirac equation under the transformations given by eq.(1) requires that the gauge function $f$ satisfy the wave equation, eq.(16).

\subsection{Vacuum continuity equations}
Let us now apply the transformation given by eq.(1) in the vacuum continuity equations ~\cite{wida}
\begin{equation}
\vec{\nabla}\rho+\frac{1}{c^2}\frac{\partial\vec{J}}{\partial t}=0\,,
\end{equation}
\begin{equation}
\vec{\nabla}\times\vec{J}=0\,,
\end{equation}
and
\begin{equation}
\vec{\nabla}\cdot\vec{J}+\frac{\partial\rho}{\partial t}=0\,.
\end{equation}
Equations (17) - (19) are invariant under eq.(1) if
\begin{equation}
(\vec{\nabla}f)\cdot\vec{J}-\frac{\partial f}{\partial t}\rho=0\,,
\end{equation}
\begin{equation}
\rho(\vec{\nabla}f)-\frac{1}{c^2}\frac{\partial f}{\partial t}\vec{J}=0\,,
\end{equation}
\begin{equation}
(\vec{\nabla}f)\times\vec{J}=0\,.
\end{equation}
Using eq.(6), eqs. (20) and (21) yield  $\rho=-\frac{\vec{v}}{c^2}\cdot\vec{J}\,$ and $\vec{J}=-\rho\,\vec{v}$. This is in agreement with eq.(22).
Now differentiate eq.(20) partially with respect to time and take the divergence of eq.(21), and add the two resulting equations, using eqs.(17) and (19) to get
\begin{equation}
\frac{1}{c^2}\frac{\partial^2f}{\partial t^2}-\nabla^2f=0\,.
\end{equation}
Once again, the invariance of the vacuum continuity equations under eq.(1) is that the gauge function $f$ satisfies the above wave equation.

\subsection{Lorenz gauge condition}
Applying the transformation given by eq.(1) in Lorenz gauge condition
\begin{equation}
\vec{\nabla}\cdot\vec{A}+\frac{1}{c^2}\frac{\partial \varphi}{\partial t}=0\,,
\end{equation}
yields
\begin{equation}
\vec{\nabla}\cdot\left(\frac{\partial f}{\partial t}\vec{A}\right)+\left(\vec{\nabla}f\right)\cdot\frac{\partial\vec{A}}{\partial t}-\frac{1}{c^2}\frac{\partial^2f}{\partial t^2}\varphi=0\,,
\end{equation}
For a vanishing field $\vec{A}\,'=0$, eq.(12) yields
\begin{equation}
-\vec{\nabla}f=\vec{A}\,,\qquad \frac{\partial f}{\partial t}=\varphi\,.
\end{equation}
The above choice is consistent with eq.(24) if $f$ satisfies the wave equation in eq.(23).

Applying eq.(24) and (26) in eq.(25) yields
\begin{equation}
\vec{\nabla}\cdot(-\varphi\, \vec{A})+\frac{\partial}{\partial t}\frac{1}{2c^2}\left(\varphi^2+c^2A^2\right)=0\,.
\end{equation}
This can be cast in an energy conservation equation form
\begin{equation}
\frac{\partial u}{\partial t}+\vec{\nabla}\cdot\vec{S}=0\,,
\end{equation}
where $\vec{S}=-\varphi\, \vec{A}$ and $u=\frac{1}{2c^2}\left(\varphi^2+c^2A^2\right)$, are the energy flux and energy density, respectively. As can be seen from eqs.(6) and (26), the vector field $\vec{A}$ is parallel to the direction of motion. Therefore, eq.(27) represents the propagation of a longitudinal wave.

We recently obtained such an equation from the generalized gauge transformations ~\cite{arbab2}.
Unlike the electromagnetic energy equation, where the \emph{r.h.s} of eq.(26) embodies the rate of doing mechanical energy $(-\vec{J}\cdot\vec{E}$), the  fields $\vec{A}$ and $\varphi$ are not associated with mechanical energy. Consequently, the wave associated with them has a great penetration power. In as much as, the energy propagates along the direction of the field, they wave associated with them will be longitudinal wave. As noted by Bass and Sch\"{o}dinger ~\cite{schrod} and lately by Aharonov and Bohm ~\cite{ahar}, the fields $\vec{A}$ and $\varphi$ are more genuine than they were dealt with in the classical electrodynamics formulation. These fields are real, since they have energy density and energy flow.

\section{Post-Galilean transformations}
We call here the  Lorentz transformations for moderately low speeds, the  \emph{post-Galilean transformations}. In his case we have
\begin{equation}
\vec{\nabla}\,'=\vec{\nabla}+\frac{\vec{v}}{c^2}\frac{\partial}{\partial t}\,,\qquad \frac{\partial}{\partial t}\,'=\frac{\partial}{\partial t}+\vec{v}\cdot\vec{\nabla}\,.
\end{equation}

Apply eq.(1) in eqs.(17) and eq.(19) to get
\begin{equation}
\vec{v}\cdot(\vec{\nabla}\rho+\frac{1}{c^2}\frac{\partial\vec{J}}{\partial t})+(\vec{\nabla}\cdot\vec{J}+\frac{\partial\rho}{\partial t})=0\,,
\end{equation}
and
\begin{equation}
(\vec{\nabla}\rho+\frac{1}{c^2}\frac{\partial\vec{J}}{\partial t})+\frac{\vec{v}}{c^2}\,(\vec{\nabla}\cdot\vec{J}+\frac{\partial\rho}{\partial t})=0\,.
\end{equation}
Equations (30) and (31) are consistent with eqs.(17) and (19). However, eq.(18) yields
\begin{equation}
\frac{\vec{v}}{c^2}\times\frac{\partial\vec{v}}{\partial t}=0\,,\qquad \vec{J}=\rho\,\vec{v}\,.
\end{equation}
We have shown recently that one can define the vorticity ($\vec{\omega}$) for a fluid by ~\cite{arbab3}
 \begin{equation}
\vec{\omega}=\frac{\vec{v}}{c^2}\times\frac{\partial(-\vec{v})}{\partial t}\,,
\end{equation}
and hence eq.(32) implies that the vacuum current is vortex free (irrotational). Hence, the vacuum current flows radially. Moreover, if one compares eq.(32) with eq.(7), one finds that the longitudinal field $\vec{E}_l=-\frac{\partial \vec{v}}{\partial t}$. Equation (32) suggests that   $\vec{E}_l$ is parallel to the velocity, $\vec{v}$. Since the vacuum electric field is also parallel to the velocity, $\vec{v}$, then the  magnetic field vanishes, as can be seen from eq.(7). Hence, the longitudinal wave associated with the vacuum has no magnetic component.

It is interesting to know that  eq.(29) is a special case of eq.(1). It corresponds to the case when, $ \vec{\nabla}f=\frac{\vec{v}}{c^2}\frac{\partial }{\partial t}\,,$ and $\frac{\partial f}{\partial t}=-\vec{v}\cdot\vec{\nabla}$.

The invariance of eq.(14) under eq.(29) yields
\begin{equation}
\vec{\nabla}\psi+\frac{\vec{\alpha}}{c}\,\frac{\partial\psi}{\partial t}=0\,\qquad\Rightarrow\qquad \frac{1}{c}\frac{\partial\psi}{\partial t}+\vec{\alpha}\cdot\vec{\nabla}\psi=0\,.
\end{equation}
This is the Dirac equation of massless particle, as evident from eq.(14). Thus, the effect of the transformations in eq.(29) is to make the mass of the particle appear to be zero. Hence, eq.(34) can be interpreted as transforming the mass, where it is zero in the boost frame but nonzero in the rest frame. Hence, the mass becomes like the electric and magnetic fields where they are zero in the boost frame while nonzero in the rest frame, see eq.(7). In this sense the mass can be defined as part of a tensor quantity.

Taking the gradient of eq.(34) using eq.(14), yields
\begin{equation}
\frac{1}{c^2}\frac{\partial^2\psi}{\partial t^2}-\nabla^2\psi+\frac{i\,m\beta}{\hbar}\,\frac{\partial\psi}{\partial t}=0\,.
\end{equation}
This represents a  wave equation that is similar to the magnetic (electric) field equation in an un-charged  conducting medium, where $\mu_0\sigma=\frac{im\beta}{\hbar}$, i.e., $\sigma=\pm\frac{im}{\mu_0\hbar}$~\cite{girr}. Equation (35) is also known as telegraph's equation. The spinor $\psi$ can be written as a two-component  column as,
 $\psi\equiv\left(\begin{array}{c} \psi_+ \\  \psi_- \end{array} \right)$, so that
 \begin{equation}
\frac{1}{c^2}\frac{\partial^2\psi_+}{\partial t^2}-\nabla^2\psi_++\frac{i\,m}{\hbar}\,\frac{\partial\psi_+}{\partial t}=0\,.
\end{equation}
 \begin{equation}
\frac{1}{c^2}\frac{\partial^2\psi_-}{\partial t^2}-\nabla^2\psi_--\frac{i\,m}{\hbar}\,\frac{\partial\psi_-}{\partial t}=0\,.
\end{equation}
For a plane wave solution of the form, $\psi=C\exp i(\omega t-\vec{k}\cdot\vec{r})$, $C$= constant, eq.(35) yields the dispersion relations
\begin{equation}
 \omega_{1, 2}=\mp\frac{mc^2}{2\hbar}\pm\sqrt{c^2k^2+\frac{m^2c^4}{4\hbar^2}}\,,
\end{equation}
where $\omega_1$ and $\omega_2$ are the energies for the states $\psi_+ $ and $\psi_-$, respectively. Equation (38) can be written as
\begin{equation}
E_{1, 2}= \hbar\omega_{1, 2}=\mp\frac{m}{2}c^2\pm\sqrt{c^2\hbar^2k^2+\left(\frac{m}{2}\right)^2c^4}\,.
\end{equation}
The first part in the LHS in eq.(39) represents the zero-point energy of the electron wave. Unlike the field wave, which has an infinite contribution to the vacuum energy, the matter wave contribution vanishes. Equation (39) can be seen as a shift of the energy of the system by the zero-energy, $E_0=\frac{m}{2}c^2$. In this case, the zero-energy ($\vec{p}=0$) will be identically zero. This is a surprising result that would resolve the infinite contribution of the zero energy of the vacuum.
It is interesting to see that such a wave packet is composed of two waves each with mass $m_*=\frac{m}{2}$ with energy $E_*=\sqrt{c^2p^2+m_*^2c^4}$, where $p=\hbar\, k$. Hence, eq.(39) yields, $E_{1, 2}=\mp E_0\pm E_*$, where $E_0=m_*c^2$. It is interesting that an electron is composed of a system with two states each has a mass that is half the mass of the electron. It is remarkable that eq.(3) describes a particle that while it has a mass behaves as a massless particle, as evident from eqs.(34) and (35).
Equation (38) implies that the above wave is composed of two waves (wavepacket) with a group velocity
\begin{equation}
v_g=\frac{\partial \omega}{\partial k}=\pm\frac{c}{\sqrt{1+\left(\frac{m_*c}{\hbar k}\right)^2}}\,.
\end{equation}
It is interesting to see that, $v_g<c$ as far as $m_*\ne0$, a fact that complies with special  theory of relativity.
At high energy ($k\rightarrow\infty$), $v_g\rightarrow  \pm \,c$ ($\omega=-\frac{m_*c^2}{\hbar}\pm\, c\,k$), while at low energy ($k\rightarrow 0$), $v_g\rightarrow \pm\,\frac{\hbar k}{m_*}$ ($\omega=-\frac{2m_*c^2}{\hbar}$ or $\omega=\frac{\hbar\,k^2}{2m_*}$).
Hence, the particle is described by a wavepacket moving with a group velocity equals to the speed of light (to the left and right), and at low energy with a group velocity that is equal to the particle velocity. We observe here that the group velocity is always less than the speed of light, except when $m=0$. This is different from the case of Dirac formalism, where the particle velocity is always equals to speed of light, even though the particle has a non vanishing mass. It can be argued that the electron is composed of two particle states one with mass $m_1=\frac{m}{2}$ and the other with mass $m_2=-\frac{m}{2}$, so that the total mass is zero. \& This entitles one state to be a particle and the other to be an antiparticle. Hence, one may argue that the electron is not very fundamental but is composed of two elementary particles. This mass has to do with the minimum energy of the vacuum which associates an energy of $\frac{1}{2}\, hf$. And since, $E=mc^2$, then the new elementary particle could also provide this minimum energy which is $E_0=\frac{m}{2}c^2$. If the field and matter zero-point contributions are equal then the total vacuum contribution will vanish. The above wave will find its applications when considering  the transport properties of electrons in solids. It is interesting to note that only when, $m=0$ then  $\omega=\pm\,ck$. This behavior mimics the electrons in graphene that has recently been discovered ~\cite{grap}.

\section{Conclusions}
We have dealt in this letter with the longitudinal wave resulting from the invariance of the equation of motion of the field and the matter.
We have shown that the material wave is a longitudinal wave and always associated with the motion of a massive particle.
This feature has not been clearly emphasized in de Broglie theory. De Broglie associated a wave nature with a particle motion,
but the nature of this wave has not been specified.  It is worth to mention that the magnetic (vortex) component of the longitudinal wave vanishes.
The post-Galilean invariance of Dirac equation yields an equation that makes the electron to behave as a massless particle while having a mass. This property is found recently to be prominent in graphene.

\end{document}